\documentstyle[12pt]{article}
\input{psfig}
\long\def\comment#1{}
\begin{document}
\title{Early Theories on the Distance to the Sun}

\author{Subhash C. Kak\\
Department of Electrical \& Computer Engineering\\
Louisiana State University\\
Baton Rouge, LA 70803-5901, USA\\
FAX: 504.388.5200; Email: {\tt kak@ee.lsu.edu}}

\maketitle

\begin{abstract}
{\it Pa\~{n}cavi\d{m}\'{s}a
Br\={a}hma\d{n}a} states that the heavens are 
1000 earth diameters, $d_e$, away
from the earth.
The sun was also taken to be halfway to the heavens, so
this suggests a distance of the sun, $R_s$, about
500 earth diameters from the earth.
The confirmation for this supposition comes from the
later theories  (c. 500 AD) from the same region and from
Greek ideas that speak of roughly the same distance.
We suggest that the original conception was $R_s \approx 500 d_e$
from which the later Indian and Greek theories diverged in 
different ways to deal with contradictory data related to outer
planet periods.

\end{abstract}

\section{Introduction}
In astronomy the story of the gradual development of the 
knowledge of the size of the solar system is a fascinating chapter.
The standard view, as presented in the well-known 
history of ancient mathematical astronomy
by Neugebauer,$^1$ is that Ptolemy in second century
AD, using a method developed by
Hipparchus, came to the conclusion that the sun is about 600
earth diameters distant from the earth.
This is the estimate which held sway during the whole of the
Middle Ages until the time of Copernicus and Brahe.
Kepler argued for a distance three times this value but it was
not before the end of the seventeenth century that it was found that
Ptolemy's estimate was wrong by a factor of about seventeen.

In this paper we sketch the early history of the knowledge
of the distance of the sun from the Indian sources,
sources that Neugebauer was not familiar with.
We now know that the knowledge of the constellations and the
planet periods can be traced at least to the
third millennium BC and the motions of the sun and the
moon to the second millennium BC.$^{2,3}$
Such knowledge must have been placed in the context of
a theory about the size of the universe.
This suggests that theories on the relative dimensions of the
solar system must be very ancient. How did the
understanding of the relative distances of the sun and the
moon emerge? And how did it evolve?

The earliest Indian evidence comes from the
{\it Rigveda} 
where there was recognition that the universe was
infinite in extent (e.g. {\it RV} 1.52.13).
Numbers as large as $10^{12}$ are described in other Vedic texts.
{\it Rigveda} 1.35.7-9 suggests that the sun is at the centre of the universe
as the rays of the sun are supposed to range from the earth to the heavens.
More practical evidence is to be found 
in texts called the Br\={a}hma\d{n}as that came to
be written as the earliest commentaries on the Vedic texts.
For example, {\it \'{S}atapatha Br\={a}hma\d{n}a
(\'{S}B)} 6.1.10 to 6.2.4 gives us a
brief account of the creation of the universe
where several elements related to the physical 
and the psychological worlds are intertwined.
Within this account the description of the
physical world is quite clear.
It begins with the image of a cosmic egg, whose
shell is the earth (6.1.11). From another cosmic egg
arises the sun and the shell of this second egg is the sky (6.2.3).
The point of this story is to suggest that the universe was perceived
at this point in the shape of an egg with the earth as the centre and the
sun going around it below the heavens.

The stars were seen to lie at varying distances with the
polestar as the furthest.

{\it Atharvaveda} 10.7 presents an image of the 
frame of the universe as a cosmic pillar ({\it skambha}).
In this the earth is taken to correspond to the base (10.7.32), the space
to the middle parts, and the heavens to the head. The sun,
in particular, is compared to the eye (10.7.33).
But there is no evidence that this analogy is to be taken in a literal
fashion.
One can be certain that in the Vedic period, the sun
was taken to be less distant than the heavens. 

It was also a common supposition in the ancient world to take the
motions of all the heavenly bodies to be uniform.
For example, such
a system of circular motions
is considered in {\it Ved\={a}\.{n}ga Jyoti\d{s}a} of 1350 BC.$^4$

The relative
distance of a body from the earth was, therefore, determined by
its period. This set up the following arrangement
for the luminaries:

{\it Moon, Mercury, Venus, Sun, Mars, Jupiter, Saturn}

Since the sun was halfway in this arrangement, it is reasonable
to assume that the distance to the sun was taken to be half
of the distance to the heavens.
The notion of the halfway distance must date from a period when
the actual periods were not precisely known or when
all the implications of the period values for the size of the
universe were not understood.
It is not clear that a purely geocentric model was visualized.
It appears that the planets were taken to go around the sun
which, in turn, went around the earth.
One evidence is the order of the planets in the days of the week
where one sees an interleaving of the planets based on the
distance from the sun and the earth, respectively; this suggests
that two points of focus, the earth and the sun, were used in 
the scheme. Further evidence comes from the fact that the planet
periods are given with respect to the sun in later texts
such as the one by \={A}ryabha\d{t}a.
It appears that the purely geocentric model may have been a later
innovation.

\section{The Pa\~{n}cavi\d{m}\'{s}a Br\={a}hma\d{n}a (PB)}
The 
{\it Pa\~{n}cavi\d{m}\'{s}a Br\={a}hma\d{n}a (PB)}
(The Br\={a}hma\d{n}a of Twenty-five Chapters)
25.10 has an account of a journey to the source of the
river Sarasvat\={\i} from the point it gets lost in the desert.
The drying up of Sarasvat\={\i} is believed to have taken place
in around 1900 BC so the text is definitely later than that epoch.
Internal astronomical evidence of the Br\={a}hma\d{n}as indicates that these 
texts date from different times in the second millennium BC.
Further evidence for this dating comes from the fact that the
Br\={a}hma\d{n}as describe rites where the interval from
the winter solstice to the summer solstice is exactly 180 days.$^5$
This makes it impossible for the rites described in
these texts to be later than the
second millennium B.C.

{\it PB} is essentially a book that deals with various rites of
different durations and so
the astronomy given in it is very incidental. The rites themselves
appear to have an astronomical intent as given by their durations:
1 through 40 days (excepting 12), 49, 61, 100, and 1000 days;
1, 3, 12, 36, 100, and 1000 years.
The
rites provide a plan for marking different portions of
the year and also suggest longer periods
of unknown meaning.

In {\it PB} 16.8.6 we have
a statement about the distance of the sun from the earth:

\begin{quote}
{\it y\={a}vad vai sahasra\d{m} g\={a}va 
uttar\={a}dhar\={a} ity \={a}hus t\={a}vad asm\={a}t
lok\={a}t svargo loka\d{h}}

The world of heaven is as far removed from this world, they say, as
a thousand earths stacked one above the other.
\end{quote}

It should be pointed out that Caland$^6$ translates this as
``...as a thousand cows standing the one above the other.'' 
Presumably, this is
because the Sanskrit word {\it gau\d{h}} has several meanings including
the primary meanings of ``earth'' and ``cow'' but considering the
context the translation by Caland is definitely wrong. Looking at the
earliest Indian book on etymology, Y\={a}ska's {\it Nirukta} which is
prior to 500 BC,$^7$
the meaning of {\it gau\d{h}}, of which {\it g\={a}va\d{h}}
is plural, is given as: ``[It] is a synonym of `earth' because it 
is extended very far, or because people go over it... It is also a 
synonym of an animal (cow) from the same root.'' ({\it Nirukta} 2.5)

Now the question arises where was the sun conceived to be in
relation to the heavens.
{\it \'{S}atapatha}, whose 
astronomy has been described elsewhere,$^{8,9}$
calls the sun the lotus of the heavens
in {\it \'{S}B} 4.1.5.17.

Let $R_s$ represent the distance between the earth and the sun,
$R_m$ be the distance between the earth and the moon,
$d_s$ be the diameter of the sun, $d_m$ be the diameter of the
moon, and $d_e$ be the diameter of the earth.

According to {\it PB}, $R_s < 1000~d_e$, and we take that

\[R_s \approx 500 d_e\]

Elsewhere,$^{10}$ we have discussed the evidence that the ancients were
aware of the relationship:

\[R_s \approx 108 d_s\]

and

\[R_m \approx 108 d_m\]

This could have been easily determined by taking a pole 
and removing it to a distance 108 times its height
to confirm that its angular size was equal to that of the sun
or the moon.
This also implies that the heavens were taken to be 216 solar
diameters from the earth.

Considering  a uniform  speed  of the sun and the moon and
noting that
the sun completes a circuit in 365.24 days and the 
moon 12 circuits in 354.37 days, we find that 

\[R_m \approx \frac{354.37 \times 500}{365.24 \times 12} d_e \]

or

\[R_m \approx 40 d_e\]

Also we have a relationship on relative sizes because

\[R_s \approx 108 d_s \approx 500 d_e\]

This means that $d_s \approx 4.63 \times d_e$.

A theory on the actual diameters of the
sun, the moon, and the earth indicates a knowledge of eclipses. The
much older {\it Rigveda} (5.40)
speaks of a prediction of the duration of a solar eclipse,
so relative fixing of the diameters of the earth, the moon,
and the sun should not come as a surprise.

Also note that the long
periods of Jupiter and Saturn 
require that the sun be much closer to the earth than the
midpoint to the heavens, or push the distance
of the heavens beyond the $1000 d_e$ of {\it PB} and perhaps also
make the distance of the sun somewhat less than $500 d_e$.
We do see these different modifications in the models from
later periods.

{\it PB} 25.10.16 also states the duration from the earth to heaven is as
long as a journey of
44 days and this is equated,
symbolically, to the travel, on horseback, between the
point where Sarasvat\={\i} is lost in the desert and its source in the
mountains. But we are not certain of the 
astronomical significance of this duration
of 44 days.

There is an interesting altar design in the
{\it \'{S}atapatha Br\={a}hma\d{n}a} (8.5) which 
represents the orbit of the sun around the earth (Figure 1).
Note that the number of bricks in the four quarters of
the year are not identical. This suggests a recognition of
the fact that the orbit of the sun was asymmetric.

\begin{figure}
\hspace*{0.5in}\centering{
\psfig{file=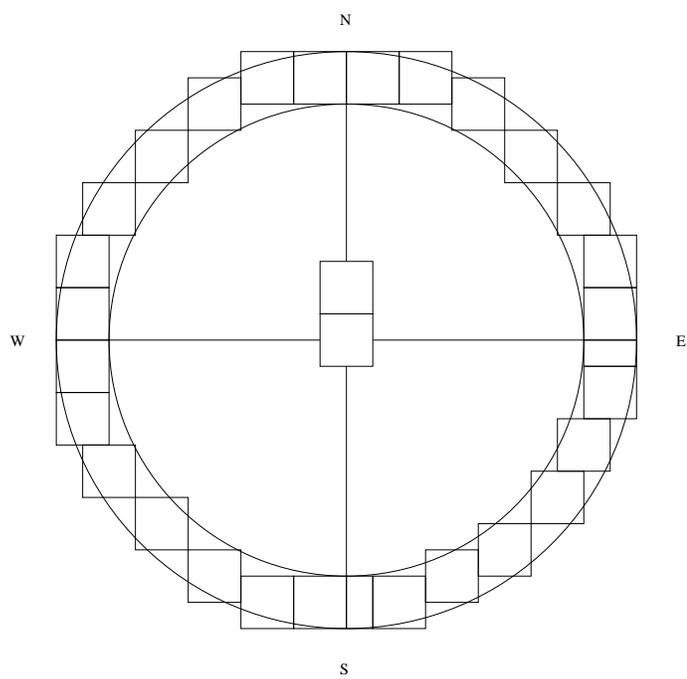,width=9cm}}
\caption{The earth's asymmetric orbit shown in an ancient 2nd millennium altar}
\end{figure}
 
\section{Planet Sizes in \={A}ryabha\d{t}a's Astronomy}

By way of comparison, we provide the values for various sizes
and distances
in a later text from India, namely
{\it \={A}ryabha\d{t}\={\i}ya (AA)} of \={A}ryabha\d{t}a$^{11}$
that dates to
c. 500 AD.
\={A}ryabha\d{t}a explains the motion of the stars as a result
of the rotation of the earth
and the motions of the planets are explained in terms of 
epicycles that, in contrast to the Greek theory, expand and
contract rhythmically.
Furthermore, he gives the planetary periods relative to the
sun which {\it appears} to be based on an ``underlying theory in
which the earth (and the planets) orbits the sun''.$^{12}$

The basic measure in this text is to
take 8,000 {\it n\d{r}} to be equal to a {\it yojana},
where a {\it n\d{r}} is the height of a man;
this makes a {\it yojana} approximately 7.5 miles.$^{13}$
{\it AA} 1.7
gives the following measures for the diameters (all in yojanas):

\vspace{0.3in}
\begin{tabular}{||l|r||} \hline
Earth ($d_e$) & 1,050.00 \\
Sun  ($d_s$)& 4,410.00 \\
Moon  ($d_m$)& 315.00 \\
Mars & 12.60\\
Mercury & 21.00 \\
Jupiter & 31.50 \\
Venus & 63.00 \\
Saturn & 15.75 \\ \hline
\end{tabular}

\vspace{0.2in}
Furthermore, {\it AA} 1.6 gives
the distance of the sun, $R_s$, 
to be 459,585 yojanas, and that of the moon, $R_m$, as 34,377 yojanas.

It follows then that in {\it AA},

\[R_s = 437.7 d_e\]

and

\[R_m = 32.74 d_e\]

Also,

\[R_s \approx 104.21 d_s\]

and

\[R_m \approx 109.13 d_m\]

Comparing the earlier figures of the {\it PB} era it is clear that
the $R_m$ had to be reduced to account for the extra time spent
in the epicyclic motions of {\it AA}.
\section{Concluding Remarks}
We first note that the idea that the sun is roughly 500 or so 
earth diameters away from us is much more ancient that Ptolemy.
So Neugebauer was wrong on two counts: first, he did not
know of any Indian connections although he
admitted$^{13}$  that the ``study of Hindu astronomy is still at its
beginning''; second, he did not recognize that the
tradition regarding the distance of the sun might be much older
in Greece itself.
This greater antiquity is in accordance with the ideas
of van der Waerden,$^{14}$
who ascribes a primitive epicycle theory to the Pythagoreans.
But it is more likely that the epicycle theory is itself much older than
the Pythagoreans and it is from this earlier source that the later
Greek and Indian modifications to this theory emerged
which explains why the Greek and the Indian models differ in
crucial details.$^{15}$

Did the idea that $R_s \approx 500 d_e$
originate at about the time of
{\it PB}, that is from the second millennium BC, or is it
older?
Since this notion is in conflict with the data on the periods of
the outer planets, it should predate that knowledge.
If it is accepted that the planet periods were known by the
end of the third millennium BC, then this 
knowledge must be assigned an even earlier epoch.
Its appearance in {\it PB},
a book dealing primarily
with ritual, must be explained as a remembrance of an old 
idea.
We do know that {\it PB} repeats, almost verbatim, the Rigvedic account of
a total solar eclipse.

Once the conflict between the planet period information and
the supposition that the heavens were 1000 earth diameters away
became clear, this supposition was dropped. Presumably,  the
theory that $R_s \approx 500 d_e$ was too entrenched by this time
and it became the basis from which different Greek and later Indian 
models emerged.
As mentioned before, Ptolemy considers an $R_s$  equal to $600 d_e$, whereas
\={A}ryabha\d{t}a assumes it to be about $438 d_e$.
Thus the Greek and the later Indian modifications to the basic idea
proceeded somewhat differently.

The ideas regarding the distance of the sun hardly changed
until the modern times.
The contradictions in the assumption that the luminaries move with
uniform mean speed and the requirements imposed by the assumed
size of the solar system led to a gradual enlargement of the
models of the universe from about twice that of the distance of the sun in {\it PB}
to one $4.32 \times 10^{6}$ times the distance of the sun by the time of 
\={A}ryabha\d{t}a.
This inflationary model of the universe in
{\it AA} makes a distinction between the distance of the sky (edge
of the universe) and
that of the stars which is taken to be a much smaller sixty times the
distance of the sun.
``Beyond the visible universe illuminated
by the sun and limited by the sky is the infinite
invisible universe'' this is stated in a commentary on {\it AA}
by Bh\={a}skara I writing in 629 AD.$^{16}$
The Pur\={a}\d{n}ic literature from India,
part of which
is contemporaneous with
\={A}ryabha\d{t}a, reconciles the finite estimates 
of the
visible universe with
the old Rigvedic
notion of an infinite universe  by postulating the existence of 
an infinite number of universes.

It is possible that the original notion that the heavens are $1000 d_e$ away
from the earth arose as a metaphor for the large extent of the
universe, given that a thousand represents a very great size in
Indo-European languages. But it is more likely that some
measurements and a theory were at the basis of this
supposition.

\section*{Notes}
\begin{enumerate}

\item Neugebauer, O., {\it A History of Ancient Mathematical
Astronomy}. Springer-Verlag, Berlin, 1975.

\item  Kak, S.C., 
"The astronomical of the age of geometric altars", 
{\em Quarterly Journal of the Royal Astronomical Society,} 36, 385-396, 1995.

\item  Kak, S.C., 
"Knowledge of the planets in the third millennium BC", 
{\em Quarterly Journal of the Royal Astronomical Society,} 37, 709-715, 1996.

\item Sastry, T.S.K., {\em Ved\={a}\.{n}ga Jyoti\d{s}a of Lagadha.}
Indian National Science Academy, New Delhi, 1985;
that the orbits are considered circular is implied by the modular
arithmetic that is used to predict the positions (see Note 2).

\item 
Kak, S.C., ``The sun's orbit in the Br\={a}hma\d{n}as,''
{\em Indian Journal of History of Science,} in press;

see also Sengupta, P.C., {\it Ancient Indian Chronology.}
University of Calcutta, Calcutta, 1947.

\item Caland, W., {\it Pa\~{n}cavi\d{m}\'{s}a Br\={a}hma\d{n}a.} 
The Asiatic Society, Calcutta, 1982, page 440.

\item Sarup, L., {\it The Nigha\d{n}\d{t}u and the Nirukta.}
Motilal Banarsidass, Delhi, 1984.

\item Kak, S.C., 
"Astronomy of the Vedic Altars",
{\em Vistas in Astronomy,} 36, 117-140, 1993.

Kak, S.C., ``The astronomy of the \'{S}atapatha Br\={a}hma\d{n}a,''
{\em Indian Journal of History of Science,} 28, 15-34, 1993.


\item  Kak, S.C., {\em The Astronomical Code of the \d{R}gveda}. 
Aditya, New Delhi, 1994.

\item See above.

\item Shukla, K.S., {\it
\={A}ryabha\d{t}\={\i}ya of \={A}ryabha\d{t}a.}
Indian National Science Academy, New Delhi, 1976.

\item Thurston, H., {\it Early Astronomy}.
Springer-Verlag, New York, 1994, page 188.

\item Neugebauer, 1975, page 7.

\item van der Waerden, B.L., {\it J. for the History of Astronomy,} 5,
175-185, 1974.

\item Burgess, E., {\it The S\={u}rya Siddh\={a}nta.}
Motilal Banarsidass, Delhi, 1989 (1860), pages 389-390.


\item Shukla, 1976, page 14.





\end{enumerate}
\end{document}